# Can providing feedback on gaze and mental-effort synchrony improve pair programming performance?

Anahita Golrang, Kshitij Sharma

Department of Computer Science, Norwegian University of Science and Technology
Trondheim, Norway

**ABSTRACT**

Pair programming is a widely used collaborative learning practice in computer science education, yet its effectiveness varies substantially due to breakdowns in coordination, attention, and cognitive regulation between partners. Drawing on theories of Socially Shared Regulation of Learning (SSRL), this paper investigates whether AI-supported feedback grounded in joint visual attention (JVA) and joint mental-effort (JME) can improve collaborative programming performance, and how feedback timing shapes learner–AI interaction. We report two experimental studies using dual eye-tracking to capture real-time indicators of collaborative regulation during debugging tasks. Study 1 examines reactive feedback that intervenes when observed JVA or JME deviates beyond predefined thresholds, while Study 2 evaluates proactive feedback that forecasts future regulatory breakdowns using machine-learning models and intervenes pre-emptively. Across both studies, feedback effectiveness is assessed through debugging success, time on task, and feedback uptake reflected in code changes.

Results show that multimodal SSRL-based feedback significantly improves collaborative performance compared to no-feedback conditions. Reactive feedback yields strong gains in debugging success and efficiency, particularly when JVA- and JME-based feedback are combined. Proactive, forecast-based feedback further enhances performance, reduces time on task, and increases constructive feedback uptake while relying less on intrusive interventions. Comparative analyses reveal that proactive feedback better preserves learner agency by maintaining optimal collaboration states and reducing escalation to disruptive support, especially for high-performing pairs. These findings demonstrate that gaze and mental-effort synchrony can serve as reliable, actionable triggers for AI-supported SSRL. The paper contributes empirical evidence and design insights for human–AI collaborative learning systems, highlighting the importance of feedback timing, transparency, and anticipatory regulation in supporting effective pair programming.

## 1. Introduction

### *1.1. SSRL and Pair programming*

Pair programming is recognized as a complex collaborative learning approach in computer science education, requiring students to work together on challenging tasks to improve computational thinking and real-world problem-solving [1]. Despite its potential benefits, the effectiveness of pair programming can be highly variable due to factors like individual programming ability, social dynamics, and group configurations [2]. To mitigate these inconsistencies, researchers have explored incorporating scaffolding mechanisms, such as integrating Large Language Models, to bolster both the efficiency and outcomes of collaborative programming by evolving team structures to include AI [3]. However, the effective utilization of AI in supporting the multifaceted areas and phases of socially shared regulation of learning in this context remains an underexplored challenge [4]. Thus, understanding the intricate relationships between individual self-efficacy, fear of failure, and the adoption of generative AI tools in programming is crucial for optimizing collaborative learning [5]. This includes investigating how AI can facilitate goal setting, strategy selection, and reflection within socially regulated learning frameworks, particularly when students encounter programming difficulties [6]. SSRL mechanisms can influence the efficacy and outcomes of students engaged in pair programming activities, focusing on their collective metacognitive processes and achievement [7]. Certain studies have examined how the dynamic interplay between individual self-regulation, co-regulation, and socially shared regulation of learning contributes to enhanced problem-solving and code quality in collaborative programming environments (e.g., [8]). Moreover, effective socially shared regulation of learning is paramount for successful collaborative programming, as it encompasses the collective negotiation of learning goals, strategies, and monitoring processes among group members [9]. This includes shared metacognitive strategies, motivational support, and emotional regulation, all of which are essential for navigating complex programming challenges in a collaborative setting [9]. Specific socially shared regulation of learning strategies, both with and without AI assistance, can impact students' collaborative problem-solving abilities and ultimately their success in pair programming tasks [10][7]. Students' confidence levels and their propensity can also rely on AI-generated code might influence the dynamics of socially shared regulation and overall programming outcomes within these collaborative environments [11]. In this contribution, we use eye-tracking in a pair programming scenario to capture the joint visual attention and joint mental-effort to provide socially shared regulatory feedback. In the first study, we provide reactive feedback (wait for the regulatory signals to go beyond a threshold); and in the second study, we provide proactive feedback (forecast the regulatory signals to go beyond a threshold).

### *1.2. SSRL and Human-AI collaboration*

Various theoretical underpinnings of socially shared regulation of learning have been adapted and expanded to encompass the unique dynamics of human-AI collaborative learning environments as well [12]. Traditional models, often rooted in human-computer interaction, are proving insufficient for grasping the complexities of human-AI partnerships, necessitating a shift towards frameworks that account for dynamic interplay, negotiation, and shared objectives [3]. Consequently, recent conceptualizations emphasize AI as a socio-cognitive teammate rather than a mere tool, acknowledging its potential for agentic participation in regulatory processes [3]. This paradigm shift reimagines AI not just as a support system but as an active co-regulator, capable of contributing to shared metacognitive strategies and adaptive feedback mechanisms within the learning process [13]. This necessitates the development of robust conceptual frameworks to understand and evaluate this emerging paradigm of human-AI interaction in learning [3]. Such frameworks move beyond considering AI as a passive tool, instead positioning it as an active participant in fostering autonomous and adaptive learning within complex AI-enhanced environments [14][3]. Specifically, the integration of artificial intelligence agents into collaborative learning environments introduces novel considerations for understanding how socially shared regulation of learning unfolds amongst learners [15]. This is particularly evident in how theoretical developments have extended traditional self-regulation concepts to include co-regulation and socially shared regulation within technology-enhanced settings, where AI can dynamically interact with metacognitive, affective, and motivational processes [16]. The integration of agentic AI transforms the AI's role from a mere instructional tool to an active participant in learning, necessitating a conceptual shift towards understanding AI as a collaborative partner capable of proactive, goal-directed actions [3]. This implies that AI agents can not only provide cognitive scaffolding through feedback but also engage in socio-emotional regulation by responding empathetically to learners' cues [13]. This expanded view aligns with the concept of self-regulated learning extending to the group level, evolving into socially shared regulation of learning, which highlights interactive planning and reflection within groups [17]. This involves not only individual self-regulation but also the collective ability of group members to co-regulate learning, thereby managing the group's learning journey effectively [18]. However, despite growing recognition of AI's potential in this domain, significant challenges remain in effectively utilizing AI to support the multifaceted areas and phases of socially shared regulation of learning [4]. To address these challenges, current research focuses on designing AI agents that can facilitate group-level regulatory processes, recognizing socially shared regulation as a critical component for successful collaborative learning [15]. In the studies included in this contribution, we designed AI-powered feedback agents that interact with learners' socially shared regulatory measurements (joint visual attention and mental-effort) to scaffold the problem solving and learning processes. We analyse their interaction with the feedback systems to understand how the feedback impacts the performance.

More specifically, in this contribution, we address the following research questions (RQ):

**Research Question 1:** How does the feedback based on the joint visual attention and the joint mental-effort measurement (reactive feedback) improves the collaborative performance?

**Research Question 2:** How does the feedback based on the joint visual attention and the joint mental-effort forecast (proactive feedback) improves the collaborative performance?

**Research Question 3:** What are the human (learner) AI interaction based differences between the reactive and proactive feedback tools?

Through these two studies, we propose the following contributions:

(1) SSRL can be reliably detected and scaffolded using multimodal indicators such as gaze and mental-effort.

(2) Proactive, forecast-based feedback represents a promising direction for AI-supported collaborative learning.

(3) Designing for learner agency requires careful attention to feedback timing, intrusiveness, and transparency.

## 2. Background and Related work

### *2.1. Learner agency, feedback and collaborative learning*

The concept of learner agency, particularly in feedback practices, denotes students' proactive engagement in seeking, providing, interpreting, and acting upon feedback [19]. This active role is crucial for fostering increased engagement, motivation, and ultimately, academic achievement [20]. Current research emphasizes that agency in feedback extends beyond self-assessment, encompassing collaborative efforts in generating and seeking meaningful feedback within social learning contexts [21][22]. When it comes to the learner agency while receiving feedback in collaborative learning , the state-of-the-art increasingly highlights a shift from students as passive recipients to active participants in an interactive and constructive process [22]. This involves learners orchestrating their own feedback generation, utilizing various resources like peer work and instructor input as comparators for self-assessment, thus promoting autonomous learning strategies [23]. Furthermore, contemporary approaches underscore the necessity of instructional scaffolding and timely feedback to cultivate reflective agency, enabling learners to actively manage and direct their learning from feedback processes ([24];[23]). Specifically, learner agency in collaborative feedback contexts is intrinsically linked to students' capacity for self- and peer-assessment, necessitating environments designed by educators to facilitate critical reflection and evaluative judgment of their own work and that of others [25]. This perspective aligns with Bandura's social cognitive theory, which posits that agency is enacted through a learner's capacity to self-reflect on their own capabilities and socially co-construct knowledge [26]. Moreover, recent investigations reveal that learner agency is not solely an individual attribute but is also profoundly shaped by contextual factors and the dynamic interplay between learners and their learning environment, including the nature of feedback interactions [19]. This shift necessitates an examination of how learners proactively regulate their learning through heightened metacognitive awareness and strategic interaction with feedback [27]. This involves creating supportive pedagogical structures that encourage learners to engage in meta-conversations about feedback's purpose and to self-regulate their learning by actively structuring their thought processes in response to feedback [28][29]. In this contribution, we consider the constructive response in the terms of changes in code while in a pair programming scenario.

### *2.2. Socially Shared Regulation of Learning (SSRL) and Design Principles*

Socially Shared Regulation of Learning (SSRL) refers to the collective processes through which learners jointly plan, monitor, and regulate cognition, motivation, and behavior while working toward a shared goal. In contrast to individual self-regulated learning, SSRL emphasizes regulation as an emergent, group-level phenomenon that unfolds through interaction, negotiation, and shared meaning making. A growing body of evidence demonstrates that successful collaboration in complex learning tasks—such as programming and debugging—depends not only on individual competence, but also on the group's ability to maintain shared awareness and regulate learning processes together.

Foundational work by Jarvela et al, [30] articulated three core design principles for supporting SSRL in computer-supported collaborative learning environments. The first principle, increasing learner awareness, highlights the importance of making otherwise invisible cognitive and regulatory states perceptible to learners. Awareness of one's own understanding, the partner's engagement, and the group's progress enables timely coordination and prevents unnoticed breakdowns in collaboration. The second principle, supporting externalization, emphasizes the role of tools and representations that allow learners to externalize internal reasoning, intentions, or difficulties, thereby making them available for joint inspection and negotiation. The third principle, prompting regulation, concerns the use of timely cues or interventions that encourage learners to initiate regulatory activities such as planning, strategy adjustment, or reflection when needed.

Empirical studies in programming education further reinforce the relevance of these principles. For instance, Leonardo Silva et al. [31] demonstrated that explicit support for regulatory strategies positively influences academic success in programming tasks, particularly when learners are encouraged to reflect on their actions and adjust strategies collectively. These findings suggest that regulation should not be treated as a purely internal or individual process, but rather as a socially mediated activity that can be scaffolded through appropriate design choices.

More recent theoretical developments extend SSRL into technology-enhanced and hybrid-intelligence contexts. [12] introduced the concept of trigger events and the Human-AI-SSRL (HASRL) model, proposing that regulatory processes can be initiated by detectable changes in learners' cognitive, behavioral, or affective states. By leveraging multimodal data and advanced analytics, such trigger-based systems open new possibilities for supporting SSRL dynamically and in real time. However, despite these advances, further work is needed to empirically explore which triggers are most informative and how design principles can be operationalized in authentic, high-cognitive-load tasks such as collaborative programming.

### *2.3. Eye-Tracking and Pair Programming*

Pair programming is one of the most widely studied collaborative practices in computer science education. Originally popularized as a core practice of Extreme Programming [32], it requires two learners to work jointly on the same codebase, continuously coordinating roles, attention, and problem-solving strategies. Empirical research consistently reports that pair programming can improve learning outcomes, problem-solving ability, and code quality, particularly in introductory programming courses [33]. From a collaborative learning perspective, Preston et al. [34] further argued that pair programming aligns closely with the defining characteristics of effective collaboration, including shared goals, mutual interdependence, and joint responsibility.

Despite these benefits, the effectiveness of pair programming is highly sensitive to coordination quality. Prior studies show that unequal participation, misaligned attention, or unbalanced cognitive effort can undermine collaboration and reduce learning gains [35, 36, 37]. These challenges are often difficult to detect in real time, as breakdowns in coordination may emerge gradually and remain implicit to the learners themselves.

Advances in eye-tracking and multimodal learning analytics have enabled more fine-grained investigation of these collaborative dynamics. Dual eye-tracking approaches allow researchers to capture how partners visually attend to shared artifacts and how their attentional patterns evolve over time. In pair programming contexts, such techniques have revealed systematic differences between successful and unsuccessful pairs, particularly in how attention is coordinated across code regions [36]. Beyond attention, pupil-based measures provide insight into learners' cognitive effort, offering a window into how mental workload is distributed and synchronized between partners.

Research combining gaze and cognitive-effort indicators suggests that effective collaboration is characterized not by constant alignment, but by a dynamic balance between convergence and divergence. Periods of shared visual focus support mutual understanding and grounding, while moments of attentional divergence enable exploration and division of labor [37]. These findings position eye-tracking not merely as an observational tool, but as a promising foundation for designing adaptive support systems that respond to emerging coordination states in collaborative programming. Together, this line of work underscores the relevance of eye-tracking as a means to operationalize otherwise implicit collaborative processes in pair programming. By linking visual attention and cognitive effort to theories of SSRL, eye-tracking provides a critical bridge between theoretical design principles and actionable, real-time support mechanisms in computer-supported collaborative learning environments.

## 3. Methodology

In this contribution, we present two studies. The first study uses a reactive feedback based on the joint visual attention, the joint mental-effort and the individual mental-efforts of the pair programmers. The second study uses a proactive feedback based on the forecasts of the joint visual attention, the joint mental-effort and the individual mental-efforts of the pair programmers.

Both the studies used exactly the same measurements from the eye-tracking (mental-effort, joint visual attention, and joint mental-effort) and the exact same measures for feedback effectiveness (debugging score, time on task). We designed five feedback tools (the reactive feedback used four out of them and the proactive feedback used all five of them).

In this section, we first present the eye tracking measurements and feedback effectiveness measures used in the two studies. Next, we present the five feedback tools designed and finally, we present the details (processes, participants and analysis) of the two studies and how do we compare them.

### *3.1. Eye-tracking Measures*

To enable real-time categorization of collaboration states, all computed indicators were measured, normalized and discretized into three categorical levels—High, Average, and Low—based on empirically validated thresholds established in prior DUET and eye-tracking studies [38, 39, 37, 40]. Categorization followed a two-standard-deviation (2SD) rule relative to each participant's resting baseline: values exceeding the baseline by more than +2SD were classified as High, values falling below –2SD were classified as Low, and measurements within the $\pm$2SD interval were designated as Average.

These thresholds represent empirically meaningful distinctions between strong, moderate, and weak synchrony. High JVA and JME values indicate effective joint focus and cognitive alignment, whereas low values reflect attentional divergence or mismatched cognitive effort. Average ME levels correspond to optimal cognitive load, while extreme values (high or low) suggest cognitive overload or disengagement, respectively.

### 3.1.1. *Joint Visual Attention (JVA)*

JVA is the amount of time the peers spend looking at the same set of objects in a given time window. The quantification of (*JVA*) is achieved by comparing the aligned attention patterns of two collaborators on a shared code document. This method overcomes issues from screen dynamics (like scrolling or terminal resizing) by establishing a persistent, non-visible **grid** based on code structure—vertically by six-codeline intervals and horizontally by screen percentiles. Raw eye-gaze coordinates are first mapped to their corresponding **codeline** by converting the gaze percentage to a pixel coordinate, which is then correlated with the known display position of the code document. The frequency of gaze occurrences in each grid slot is accumulated over a short interval (30 seconds) for both participants. Finally, the *JVA* score, which represents the degree of their shared attentional focus, is calculated using **cosine similarity** between the two resulting gaze frequency distribution grids.

### 3.1.2. *mental-effort and Joint mental-effort (JME)*

Each participant's cognitive load was quantified using the *Index of Pupillary Activity (IPA)* [38], a wavelet-based signal decomposition of the pupil diameter time series that captures high-frequency oscillations linked to mental-effort. The resulting ME signals ($ME_1$ and $ME_2$) were segmented into non-overlapping 10-second windows. Within each window, JME was computed using cosine similarity, producing a time-aligned similarity score that reflects how closely the two partners' cognitive load patterns evolved over the same interval. Higher values indicate stronger synchrony in mental-effort, whereas lower values reflect divergence or imbalance in the pair's engagement. JME, defined as cognitive similarity or the similarity in mental-effort between the two peers in the dyad, is calculated server-side from the time series of the individual ME ($ME_1$ and $ME_2$) values. To compute JME, the resulting individual ME time series are first discretised into a manageable integer range. The similarity, or synchrony, between these two discretized time-series is then calculated using **cross-recurrence quantification** [41], which yields the JME score that represents the degree to which the dyad is sharing a similar cognitive state across time. More specifically, to compute this measure, we first estimate each participant's cognitive load from the pupil dilation data using the method described by Duchowski et al. (2018). The resulting values are then discretized into integers ranging from 0 to 10. After obtaining the cognitive load time series for both members of the dyad, we calculate the cross-recurrence between them following the approach proposed by Richardson et al. (2007).

## *3.2. Feedback Effectiveness Measures*

(1) Debugging success: The total number of bugs successfully fixed in the given task time. This metric measures the effectiveness of the pair in correctly identifying and resolving errors. The score is increased by one, indicating the number of bugs solved plus the bug the participants were working on as the task ended.
(2) Debugging time on task: The total time on task is the total time spent in seconds. This metric measures the efficiency of the collaboration.
(3) Feedback uptake: This is defined, using the code snippets, by the changes in the code made by the pair after receiving a given piece of feedback and before the next feedback was triggered. This symbolises the learner agency in the two studies by computing the cautious effort made by the programmers on the artifact that they were collaborating on.

### 3.3. *Feedback Tools*

- **GitHub Copilot:** In this study we employ only GitHub Copilot's AI-driven autocomplete, a feature that generates context-aware code suggestions directly within the editor. Prior work shows that Copilot can accelerate task completion and reduce perceived cognitive effort [42] , though its influence on learning remains underexplored. Within our framework, "enabling Copilot" refers specifically to activating this autocomplete mechanism as a momentary scaffold when forecasted mental-effort patterns indicate cognitive strain. The demonstration of this feedback can be seen in Fig 1.

```
100
101        # Do we bounce off the left of the screen?
102        if self.x <= 0:
103            self.direction = (360 - self.direction) % 360
104            self.x = 1
105
106        # Do we bounce of the right side of the screen?
107        if self.x > self.screenwidth - self.width:
108            self.direction = (360 - self.direction) % 360
109            self.x = self.screenwidth - self.width - 1
110
111        # Did we fall off the bottom edge of the screen?
112        if self.y > 600:
               return True
```

**Figure 1.:** Github copilot feedback

- **Dual Text Selection** Broadcasted text selection is implemented as an always-on mechanism in which each collaborator's cursor selection is continuously shared, offering a lightweight visual cue that facilitates rapid joint focus. Modern shared editors, including Visual Studio Code's Live Share used in our experiment, inherently support dual text selection, enabling both partners to highlight code independently while preserving mutual awareness. This design aligns with the role of Joint Visual Attention (JVA)—the similarity of partners' gaze patterns—as a core indicator of coordination in concurrent collaboration. Figure 3 illustrates this feedback visualization.

```
100
101        # Do we bounce off the left of the screen?
102        if self.x <= 0:
103            self.direction = (360 - self.direction) % 360
104            self.x = 1
105
106        # Do we bounce of the right side of the screen?
107        if self.x > self.screenwidth - self.width:
108            self.direction = (360 - self.direction) % 360
109            self.x = self.screenwidth - self.width - 1
110
111        # Did we fall off the bottom edge of the screen?
112        if self.y > 500:
113            self.direction = (180 - self.direction) % 360
114
```

**Figure 2.:** Dual Text Selection feedback

- **Gaze-Awareness Tool:** A translucent, colored rectangle indicates the partner's current gaze region (approximately nine code lines).activating visual feedback only when the relevant synchrony metric falls below a predefined threshold. The visualization of this scafolding can be found in Figure **??**. This feedback is activated when the JVA is low.

**Figure 3.:** Gaze-Awareness Tool

- **Dialog Prompt:** This feedback is intentionally designed as the least disruptive form of support and is triggered when the pair's JME is low, indicating a possible asymmetry in understanding. When activated, a small unobtrusive prompt appears in the bottom-right corner of the editor (Figure 4), encouraging the partners to initiate brief dialogue. This conversational nudge is intended to strengthen shared understanding and guide the pair toward more aligned—and more optimal mental-effort levels.

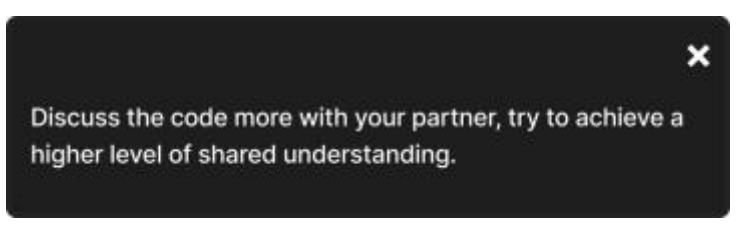

**Figure 4.:** Dialog Prompt

- **Task-Based Hint :** The task-based hint constitutes the most disruptive form of support and is therefore reserved for situations where None of the previous scaffolding have been productive and both collaborators exhibit extreme mental-effort levels, indicating substantial cognitive strain or stagnation. When triggered, the system presents a small window in which the user selects the current task and the specific bug, after which a targeted hint is provided (Figure 5). By directing the pair toward a relevant section of the code and clarifying the underlying issue, the hint is designed to realign their problem-solving trajectory and restore cognitive balance. Although more intrusive than other feedback types, this intervention is intended to produce a strong corrective effect, particularly during episodes of prolonged difficulty

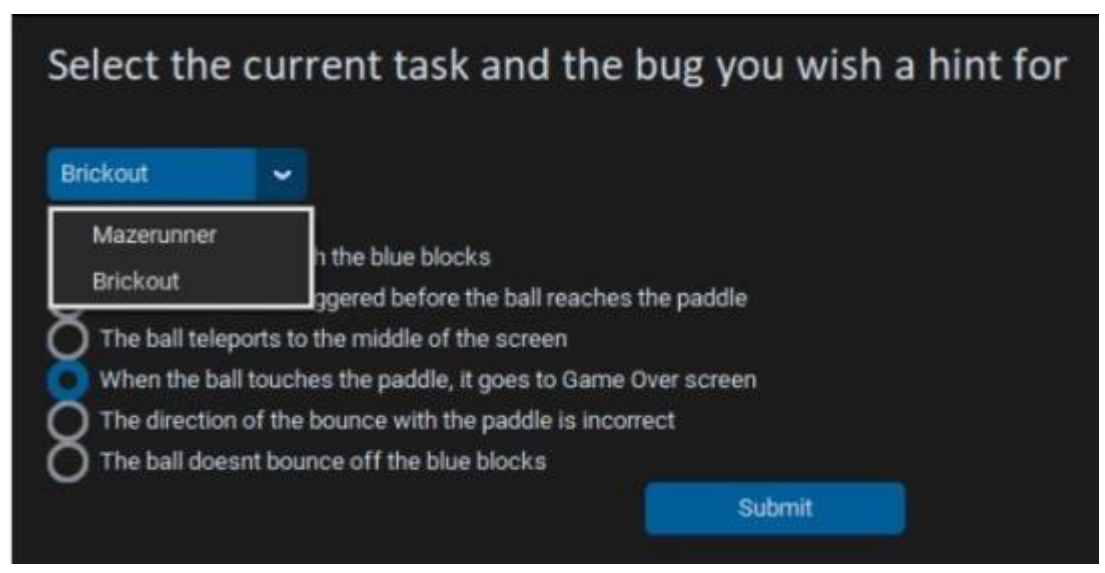

**Figure 5.:** Hint Prompt

### *3.4. Study 1: Reactive feedback*

#### *3.4.1. Participants and Procedure*

In this study, we employed a between-subject design with four conditions: 1) control (no feedback); 2) JVA feedback; 3) JME feedback; 4) Both (combined JVA and JME) feedback. In each of the conditions, we recruited 30 pairs (120 pairs in total, 240 participants, 81 females, mean age 21.4, variance age 3.2) to participate in the pair programming tasks. The pairs were randomly formed. The participants were welcomed into the laboratory as pairs, where they first signed a consent form and then received a demonstration of the feedback tool they would be using. In the demo, they were informed about the interactions with the feedback tools, including when they would be triggered, what changes would be made in the code editor, and how they could switch the feedback on and off. Then the two eye-tackers were calibrated for the individual programmer. Finally, they were shown how the correct version of the program would function (not the correct code itself) that they received to debug and were provided the bug-infused version of the code to solve. The participants were informed that there were only logical bugs in the code and no syntax errors.

#### *3.4.2. Feedback mechanism*

The reactive component responds to the immediate, moment-to-moment conditions of the pair. Feedback is triggered whenever the real-time measurements deviate more than 2 standard deviations (2SD) from each participant's resting baseline, enabling rapid correction when the system detects abrupt or pronounced shifts. The detailed scenario logic and feedback combinations are shown in Appendix A. For the combined JVA and JME condition (**both**), we used three feedback tools: offering help by enabling the gaze-awareness tool, prompting dialogue initiation, and providing a task-based hint. For the JVA-only condition, we use two feedback tools: offering help through enabling the gaze-awareness tool and a prompt to initiate dialogue. For the JME-only condition, we used only one feedback tool: providing a task-based hint. The triggers are described in Appendix A.

#### 3.4.3. Data Analysis

To answer the first research question, we conducted an ANOVA with the debugging success metrics as the dependent variable and the feedback conditions as the independent variable. We also conducted pair-wise ANOVAs to compare the conditions further. All the p-values were corrected using Bonferroni corrections for multiple comparisons. Prior to conducting ANOVA, we verified the homoscedasticity of the variables using Levene's Test; and the normality using Shapiro-Wilk test.

### 3.5. Study 2: Proactive feedback

#### 3.5.1. Participants and Procedure

In this study, we employed a within-subject design with two conditions: 1) control (no feedback) condition; 2) experimental (combined JVA and JME forecasting feedback) condition. We recruited 26 pairs to participate in the pair programming tasks. The pairs were randomly formed. The participants were welcomed into the laboratory as pairs, where they first signed a consent form and then received a demonstration of the feedback tool they would be using. In the demo, they were informed about the interactions with the feedback tools, including when they would be triggered, what changes would be made in the code editor, and how they could switch the feedback on and off. Then the two eye-tackers were calibrated for the individual programmer. Finally, they were shown how the correct version of the program would function (not the correct code itself) that they received to debug and were provided the bug-infused version of the code to solve. The participants were informed that there were only logical bugs in the code and no syntax errors.

#### 3.5.2. Feedback mechanism

The proactive component anticipates future collaboration states by relying on forecasted values of the system's key measures. An Extreme Gradient Boosting (XGBoost) model generates real-time predictions over a 30-seconds horizon, allowing the system to intervene before suboptimal states fully materialize. The feedback logic operated on collaboration states forecasted 30 seconds in advance by an XGBoost model. The model's predictions for JVA, JME, and ME were discretized into High (H), Average (AVG), and Low (L) categories and then compared against a desired collaboration matrix (see Table 9 in Appendix A). 2 The detailed scenario logic and feedback combinations are shown in Appendix A. For the feedback condition, we employed the complete logic presented in the Appendix.

#### 3.5.3. Data Analysis

To answer the second research question, we conducted a paired t-test with the debugging success metrics as the dependent variable and the feedback conditions as the independent variable. Prior to conducting t-test, we verified the homoscedasticity of the variables using Breusch-pegan Test; and the normality using Shapiro-Wilk test.

### 3.6. Comparing Reactive and Proactive feedback

To answer the third research question, we compared **"both"** condition from the first study and the **experimental** condition from the second study. For this comparison, we analysed the interaction of the programmers with the feedback tool. To this end, we employed one ANOVA for each study, with performance and feedback types as the independent variables and the amount of feedback received as the dependent variable. Then we compare the effect sizes of the common components (e.g., performance, dialogue prompts, gaze-awareness feedback, hint prompt).

## 4. Results & Findings

The pairs were given opportunities to pause the feedback system for two minutes or cancel the feedback system completely at any time during the experimental sessions. They were also given an opportunity at every instance when a feedback was triggered to press a button and ignore the feedback. None of the dyads in either of the studies chose to block the feedback anytime.

### *4.1. Study 1: Reactive Feedback*

An ANOVA with the feedback category as the independent variable and debugging success as the dependent variable shows a clear difference in the debugging success among the feedback categories ($F[3,116] = 214.87$, $p < .0001$, figure 6).

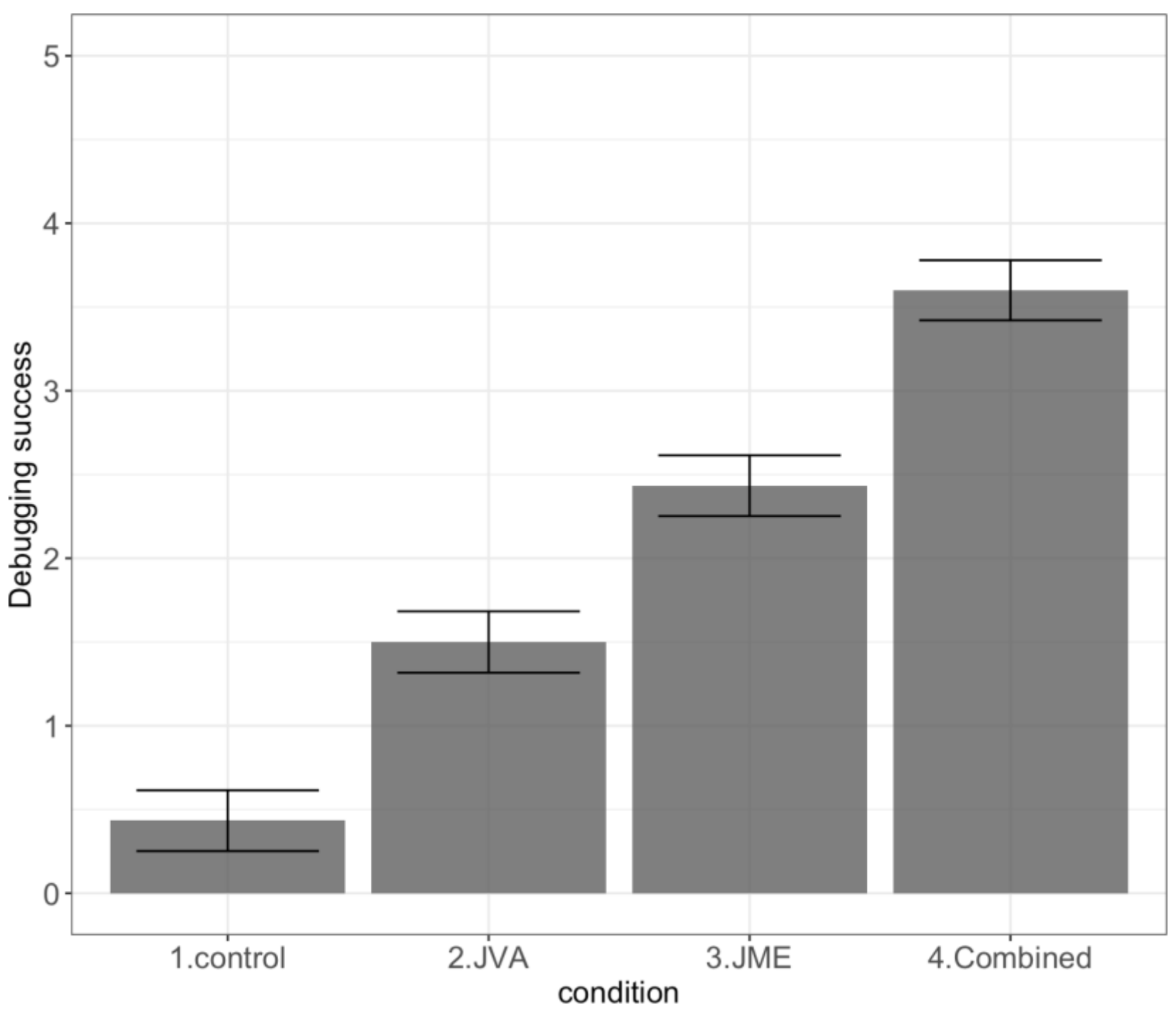

**Figure 6.:** Debugging success across the reactive feedback conditions.

Further pairwise-ANOVAs (Table 1, columns 3 and 4) show that the debugging success is the highest in the combined condition, followed by the debugging success in the JME condition, and followed by the debugging success in the JVA condition. Finally, the debugging success is the lowest for the control condition among all the conditions (Figure 6)

Another ANOVA with the feedback category as the independent variable and debugging time on task as the dependent variable shows a clear difference in the debugging time on task among the feedback categories ($F[3,116] = 42.75$, $p < .0001$, Figure 7).

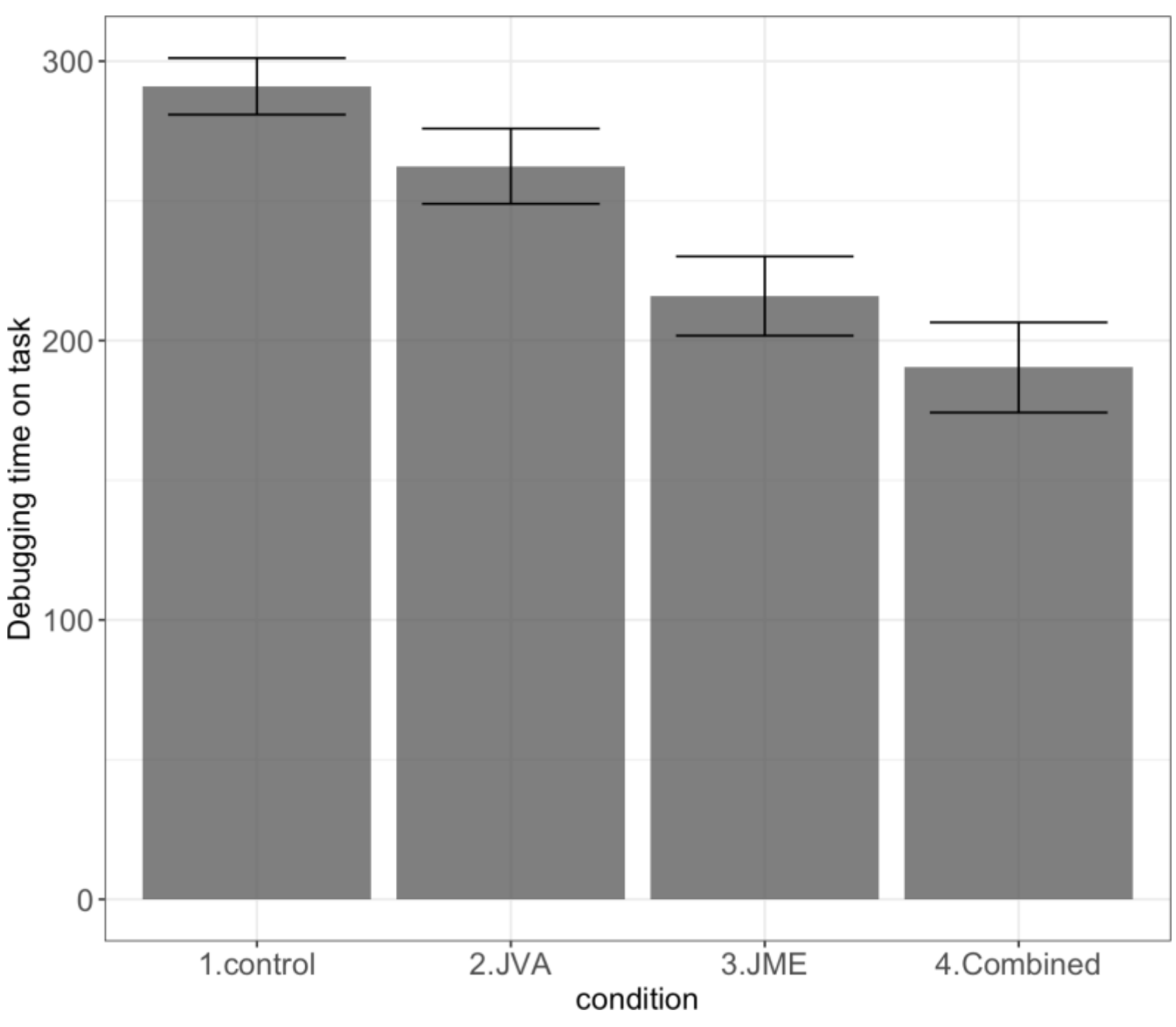

**Figure 7.:** Debugging time on task across the reactive feedback conditions.

Further pairwise-ANOVAs (Table 1, columns 5 and 6) show that the debugging time on task is the lowest in the combined condition, followed by the debugging time on task in the JME condition, and followed by the debugging time on task in the JVA condition. Finally, the debugging time on task is the highest for the control condition among all the conditions (Figure 7)

Finally, an ANOVA with the feedback category as the independent variable and feedback uptake as the dependent variable shows a clear difference in the feedback uptake among the feedback categories ($F[3,116] = 289.66$ , $p < .0001$, Figure 8).

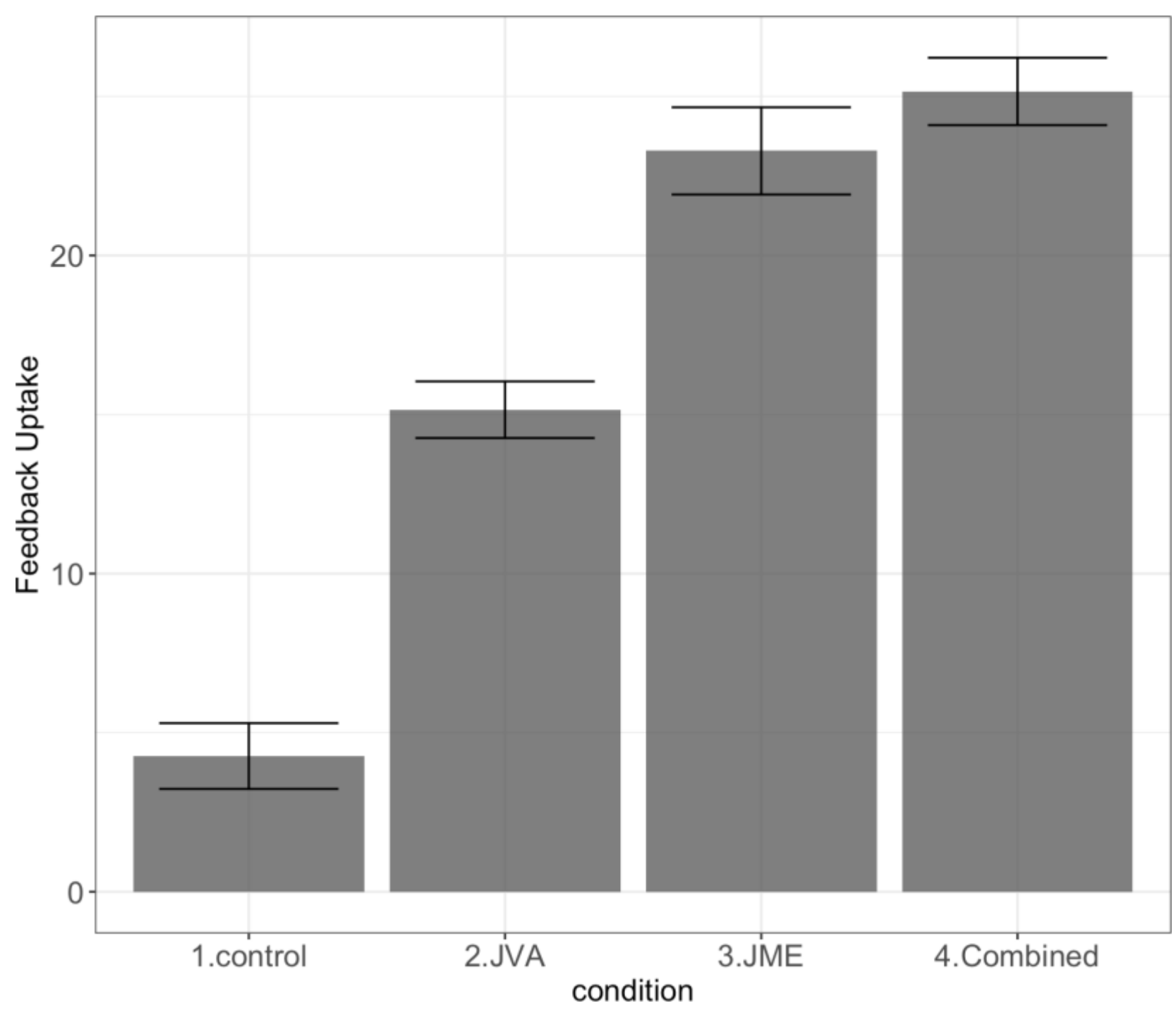

**Figure 8.:** Feedback uptake across the reactive feedback conditions.

Further pairwise-ANOVAs (Table 1, columns 7 and 8) show that the feedback uptake is the lowest in the combined condition, followed by the feedback uptake in the JME condition, and followed by the feedback uptake in the JVA condition. Finally, the feedback uptake is the highest for the control condition among all the conditions (Figure 7)

**Table 1.:** Pairwise comparisons between the reactive feedback conditions for debugging success and time on task.

| | | Pairwise comparisons | | | | | |
|---|---|---|---|---|---|---|---|
| | | Debugging Success | | Debugging time on task | | feedback uptake | |
| | | F [1, 58] | P-value | F [1, 58] | P-value | F [1, 58] | P-value |
| **Control** | **JVA** | 66.58 | <.0001 | 11.19 | .001 | 247.52 | <.0001 |
| **Control** | **JME** | 236.2 | <.0001 | 72.17 | <.0001 | 477.26 | <.0001 |
| **Control** | **Combined** | 598.91 | <.0001 | 108.48 | <.0001 | 774.44 | <.0001 |
| **JVA** | **JME** | 50.97 | <.0001 | 21.97 | <.0001 | 96.38 | <.0001 |
| **JVA** | **Combined** | 261 | <.0001 | 45.68 | <.0001 | 203.39 | <.0001 |
| **JME** | **Combined** | 81.29 | <.0001 | 5.50 | .02 | 4.53 | .03 |

### *4.2. Study 2: Proactive Feedback*

First, we analysed the order-effect on all the three measurements for feedback efficiency because this experiment was a within-subject design and we had balanced the order of the control task and the experimental task. We observed no order effect on any of the dependent variables i.e., debugging success ($t[24] = 0.45$, $p > .05$), debugging time on task ($t[24] = 0.56$, $p > .05$), and the feedback uptake ($t[24] = 1.01$, $p > .05$).

An t-test with the feedback category as the independent variable and debugging success as the dependent variable shows a clear difference in the debugging success among the feedback categories ($t[49.96] = -13.51$, $p < .0001$, figure 9). The debugging success is significantly higher in the feedback condition than the control condition.

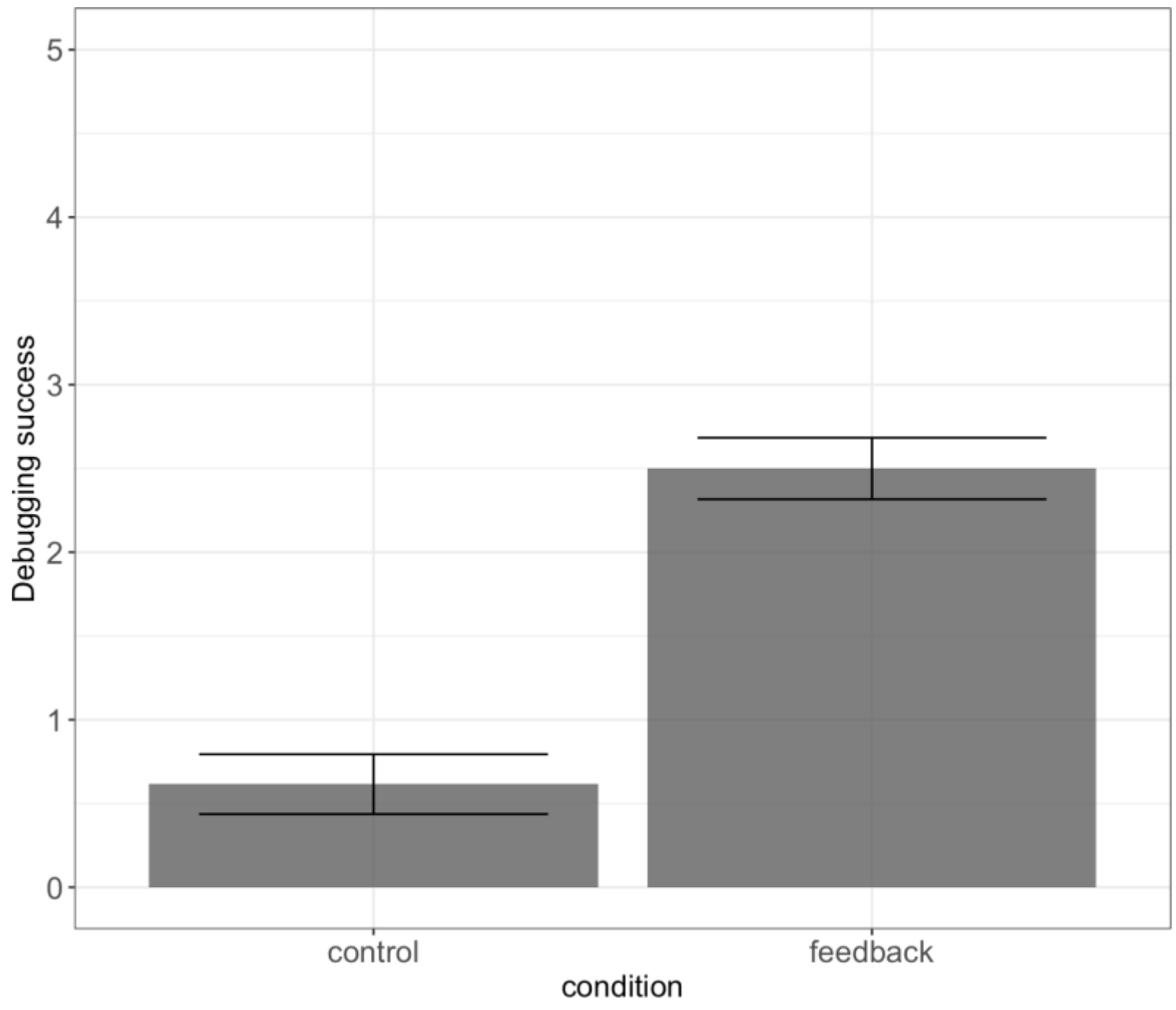

**Figure 9.:** Debugging success across the proactive feedback conditions.

An t-test with the feedback category as the independent variable and debugging time on task as the dependent variable shows a clear difference in the debugging time on task among the feedback categories ($t[44.70] = 4.39$, $p < .0001$, figure 10). The debugging time on task is significantly lower in the feedback condition than the control condition.

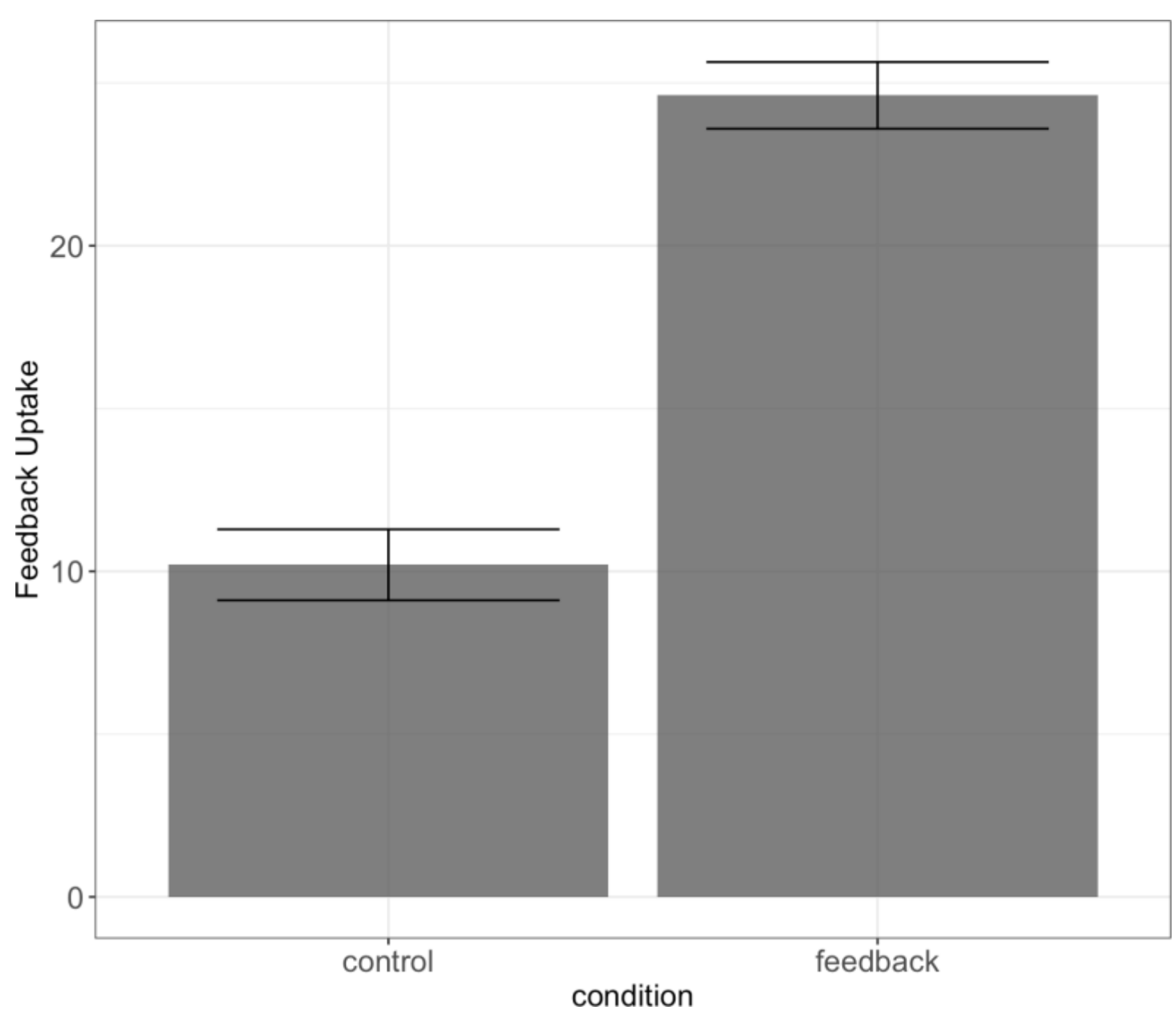

**Figure 10.:** Debugging time on task across the proactive feedback conditions.

An t-test with the feedback category as the independent variable and feedback uptake as the dependent variable shows a clear difference in the feedback uptake among the feedback categories ($F[49.81] = -17.69$, $p < .0001$, figure 11). The feedback uptake on task is significantly higher in the feedback condition than the control condition.

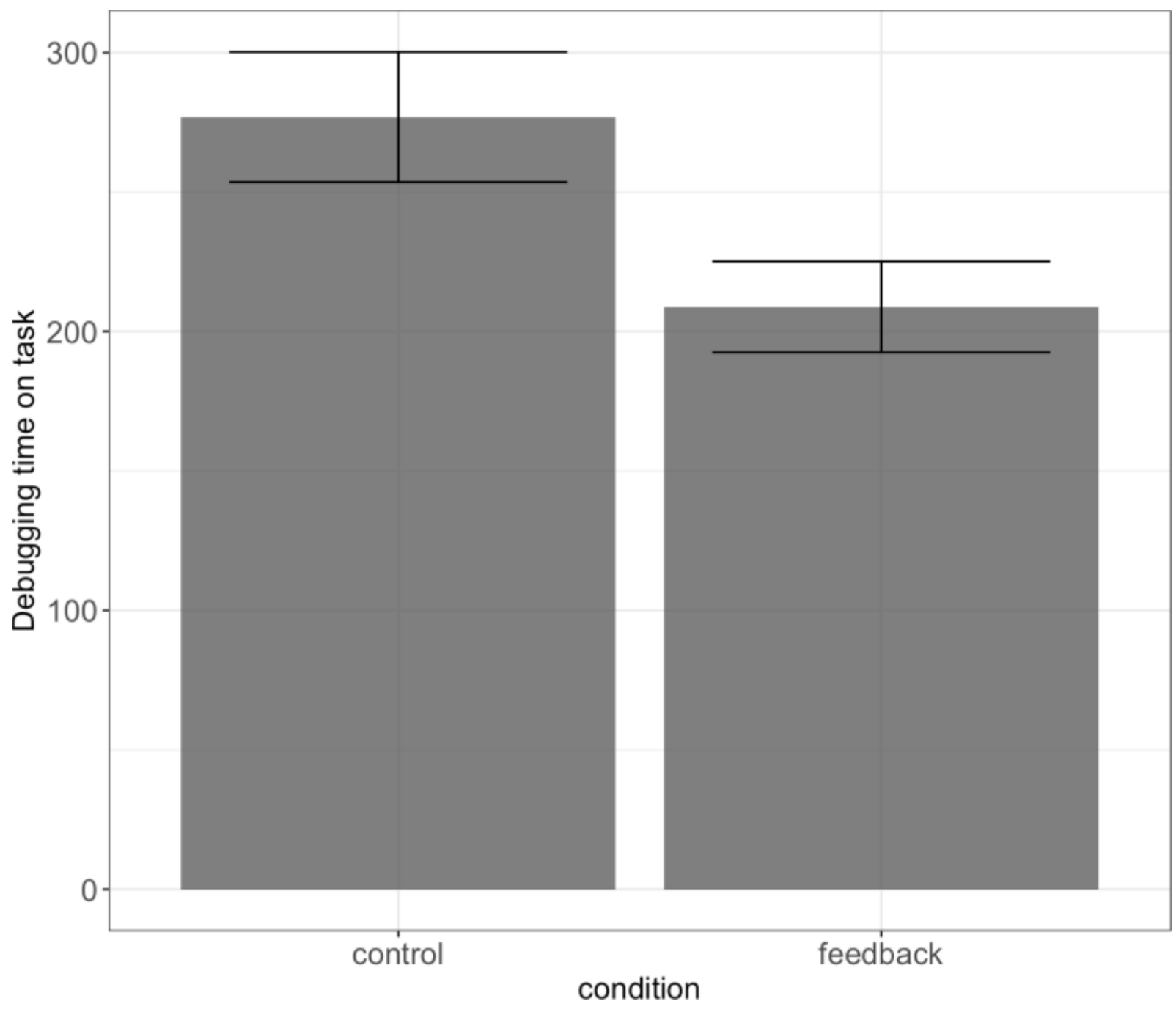

**Figure 11.:** Feedback uptake across the proactive feedback conditions.

### *4.3. Reactive vs Proactive Feedback*

To compare the reactive and proactive feedback in the two studies, we compare the feedback that the students get and how did they performed within the feedback condition. In the proactive feedback study, we had combined the feedback based on JVA and JME, therefore, we will only use the Combined condition from the reactive feedback study.

For the reactive feedback study, we observe an interaction effect of performance level (high vs low) and feedback type (dialogue prompt, gaze-aware feedback, hint prompt) on the amount of feedback provided (Table 2). Low performers get more feedback then the high performers ($F[1,28] = 23.04$, $p < .0001$, Figure 12). Dialogue prompt was the least provided feedback and there was not difference in the gaze-awareness and the hint prompts for the reactive study (Table 4, Figure 12). Finally, for the interaction effect, we observe that high performers get significantly more ”hint prompt” feedback, the low performers receive significantly more ”dialogue prompt” and ”gaze-awareness tool” (Table 4, Figure 12)

**Table 2.:** The effect of performance level (high vs low) and feedback type (dialogue prompt, gaze-aware feedback, hint prompt) on the amount of feedback provided

| | Df | F-value | p-value |
|---|---|---|---|
| **Performance** | 1 | 23.04 | <.0001 |
| **Feedback Type** | 2 | 84.10 | <.0001 |
| **Interaction (Performance : Feedback Type)** | 2 | 91.03 | <.0001 |

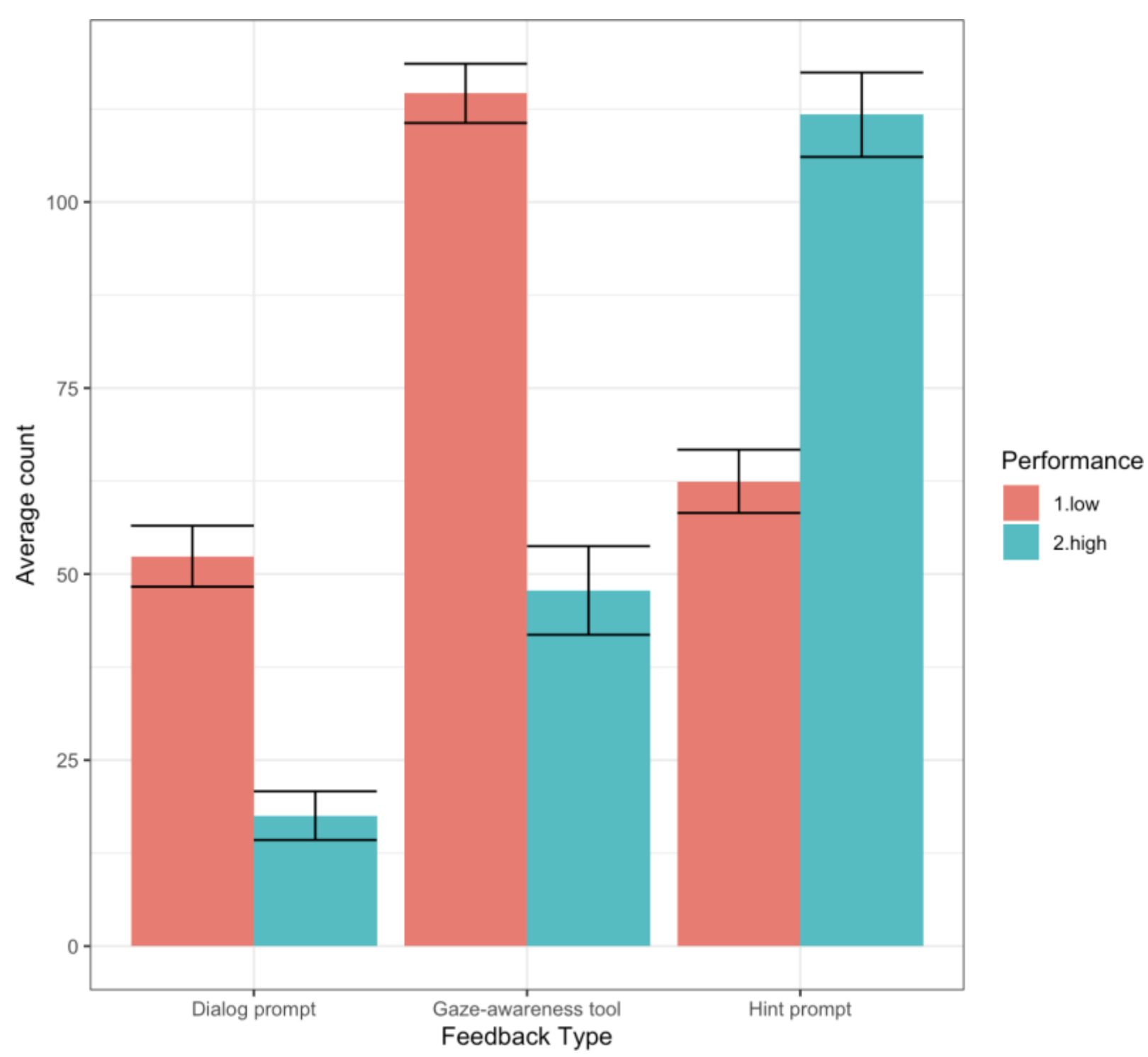

**Figure 12.:** Amount and types of feedback provided in the reactive feedback study and performance in the combined condition.

**Table 3.:** Difference between high and low performers in the reactive feedback condition for the different feedback types.

| | **Dialogue prompt** | **Gaze-awarenesss Feedback** | **Hint propmt** |
|---|---|---|---|
| **T-value** | 13.72 | 9.78 | -10.23 |
| **p-value** | <.0001 | <.0001 | <.0001 |

**Table 4.:** Pairwise comparison for the amount of feedback given the feedback types in the reactive feedback for the combined condition.

| | **Gaze-awarenesss Feedback** | **Hint propmt** |
|---|---|---|
| **Dialogue prompt** | -6.08<br><.0001 | -8.23<br><.0001 |
| **Gaze-awarenesss Feedback** | | -0.70<br>.48 |

For the proactive feedback study, we observe an interaction effect of performance level (high vs low) and feedback type (do nothing, github copilt, dialogue prompt, gaze-aware feedback, hint prompt) on the amount of feedback provided (Table 5). Low performers get more feedback then the high performers ($F[1,24] = 13.55$, $p < .0001$, Figure 13). Dialogue and hint prompts were the least provided feedback types (no significant difference in these two types), followed by gaze-awareness feedback, followed by the do nothing feedback. Finally, the github copilt was the most initiated feedback type (Table 7, Figure 13). Finally, for the interaction effect, we observe that high performers get significantly more ”do nothing” feedback, the low performers receive significantly more ”dialogue prompt”, ”github copilot” and ”gaze-awareness tool” (Table 6, Figure 13).

**Table 5.:** The effect of performance level (high vs low) and feedback type (Do nothing, Github copilot, dialogue prompt, gaze-aware feedback, hint prompt) on the amount of feedback provided

| | Df | F-value | p-value |
|---|---|---|---|
| **Performance** | 1 | 13.55 | <.0001 |
| **Feedback Type** | 4 | 104.10 | <.0001 |
| **Interaction (Performance : Feedback Type)** | 4 | 90.03 | <.0001 |

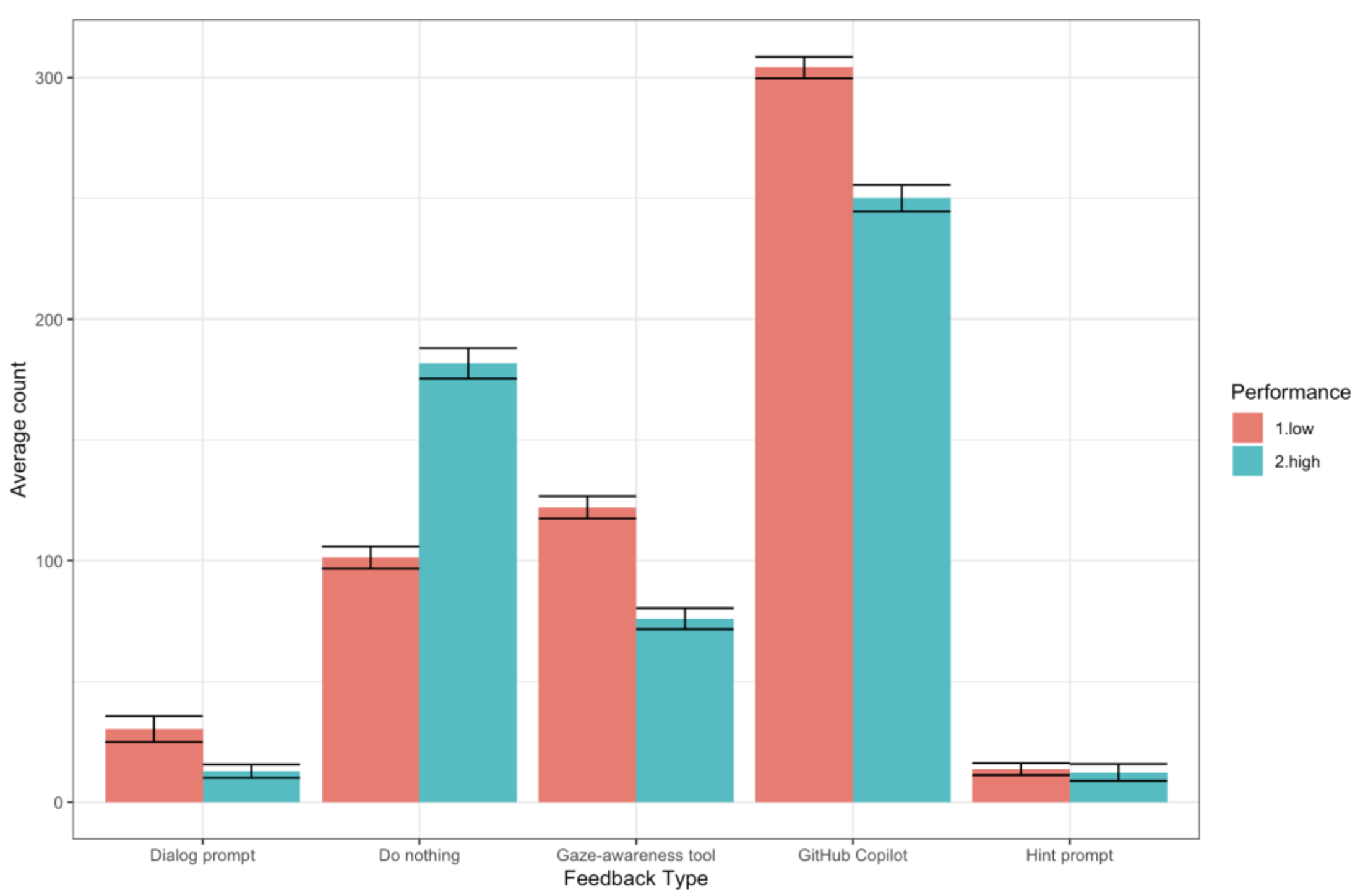

**Figure 13.:** Amount and types of feedback provided in the proactive feedback study and performance in the experimental condition.

**Table 6.:** Difference between high and low performers in the proactive feedback condition for the different feedback types.

| | Do nothing | Github copilot | Gaze awareness | Dialogue prompt | Hint prompt |
|---|---|---|---|---|---|
| **T-value** | -14.06 | 10.46 | 9.91 | 3.97 | 0.44 |
| **p-value** | <.0001 | <.0001 | <.0001 | .0009 | .66 |

**Table 7.:** Pairwise comparison for the amount of feedback given the feedback types in the proactive feedback for the experimental condition.

| | **Github Copilt** | **Dialogue prompt** | **Gaze-awarenesss Feedback** | **Hint propmt** |
|---|---|---|---|---|
| **Do nothing** | -13.04 <.0001 | -13.39 <.0001 | 4.27 .0001 | 14.85 <.0001 |
| **Github Copilt** | | -38.81 <.0001 | -22.60 <.0001 | 42.82 <.0001 |
| **Dialogue prompt** | | | -13.26 <.0001 | 2.70 .01 |
| **Gaze-awarenesss Feedback** | | | | 16.04 <.0001 |

Table 8 shows the comparison between the two studies. We observe that the effect sizes for the amount of feedback received are all larger for the reactive feedback than the proactive feedback. We observe that high performers receive significantly less feedback than the low performers in the compared conditions. However, this difference is larger in the reactive study than in the proactive study. Next, we observe that high performers receive significantly fewer dialogue prompts and significantly fewer gaze-awareness feedback than the low performers in both studies; however, both of these differences are also larger in the reactive study. Finally, we observe that in the reactive study, high performers receive significantly more hint prompts than low performers; in contrast, there is no such difference in the proactive study.

**Table 8.:** Comparison, based on the Cohen's d, between the reactive and proactive feedback tools. NS = not significant

| | **Effect sizes (Cohen's d)** | |
|---|---|---|
| **Component** | **Study 1 Reactive Feedback** | **Study 2 Proctive Feedback** |
| **Performance** | 1.26 | 0.96 |
| **Dialogue prompt** | 5.60 | 3.74 |
| **Gaze-awareness feedback** | 3.99 | 1.50 |
| **Hint prompt** | 4.71 | 0.16 (NS) |

## 5. Discussion

This study examined whether feedback grounded in joint visual attention (JVA) and joint mental-effort (JME) can improve collaborative performance in pair programming, and how reactive and proactive feedback designs differentially shape socially shared regulation of learning (SSRL) and human–AI interaction. Building on prior work emphasizing SSRL as a core mechanism underlying successful collaborative programming (e.g., [7, 6]), our findings provide converging evidence that making implicit regulatory states visible and actionable can substantially enhance collaborative outcomes.

Across two studies, we demonstrate that multimodal, AI-supported feedback systems can function as effective co-regulators of learning by supporting awareness, externalization, and timely regulation, three design principles originally articulated for CSCL environments [30] and more recently extended to Human–AI-SSRL contexts [12].

### 5.1. Reactive Feedback and the Role of Awareness in SSRL (RQ1)

The results of Study 1 show that reactive feedback based on real-time deviations in JVA and JME significantly improves debugging success and efficiency compared to a no-feedback control condition. These findings align with prior research demonstrating that breakdowns in coordination, such as misaligned attention or uneven cognitive effort, are a major source of variability in pair programming outcomes [35, 36, 37]. By detecting these breakdowns as they emerge and prompting regulatory activity, the system helped pairs maintain productive collaboration.

The superiority of the combined JVA+JME condition over single-indicator conditions supports the view that SSRL is a multi-dimensional phenomenon that cannot be reduced to attentional alignment alone or cognitive synchrony alone [17]. Prior eye-tracking studies in pair programming have shown that successful pairs exhibit dynamic patterns of convergence and divergence in gaze [36, 37]. Our results extend this work by showing that intervening on these patterns, rather than merely observing them, can directly improve performance.

From a design perspective, the reactive feedback instantiated all three SSRL design principles proposed by Järvelä et al. [30]:

(1) Increasing awareness, by visualizing partners' gaze regions when JVA was low;
(2) Supporting externalization, by prompting dialogue when JME indicated cognitive asymmetry; and
(3) Prompting regulation, by offering task-based hints during sustained cognitive strain.

The strong performance gains observed in the combined condition suggest that effective support for collaborative programming requires coordinated scaffolding across these principles, consistent with findings from computational scaffolding research in programming education [6], [31] .

Interestingly, feedback uptake, as measured by observable code changes following feedback, was lowest in the combined reactive condition, despite this condition yielding the best performance. This pattern resonates with SSRL theory, which emphasizes that effective regulation often prevents breakdowns rather than producing visible corrective actions after failure [9]. When awareness and regulation are well supported, learners may require fewer overt interventions at the artifact level, resulting in higher efficiency with less observable feedback uptake.

### 5.2. Proactive Feedback and Anticipatory Regulation (RQ2)

Study 2 demonstrates that proactive, forecast-based feedback further improves collaborative performance beyond a control condition. By predicting future declines in JVA and JME and intervening before misalignment fully materialized, the system supported what can be described as anticipatory SSRL. This finding empirically substantiates recent theoretical proposals that regulatory processes can be triggered by early indicators of change in cognitive or behavioral states, rather than only by overt failures [12].

The proactive condition yielded higher debugging success, reduced time on task, and increased feedback uptake, suggesting that learners were better able to act on feedback when it was delivered earlier and in a more preventative manner. This aligns with research on learner agency in feedback processes, which emphasizes that timely, interpretable feedback enhances learners' capacity to regulate their learning proactively rather than reactively [19, 22, 23].

Moreover, the proactive system's reliance on forecasting models positions AI not merely as a responsive tool but as an agentic collaborator capable of contributing to planning and monitoring phases of SSRL. This supports recent reconceptualizations of AI as a socio-cognitive teammate in collaborative learning environments [3, 14, 15]. By intervening before cognitive overload or attentional divergence escalated, the AI system helped sustain productive collaboration trajectories without disrupting learners' sense of control.

### *5.3. Reactive vs. Proactive Feedback and Learner Agency (RQ3)*

Comparing the reactive and proactive systems reveals important differences in how human–AI interaction unfolds under different feedback logics. In both studies, low-performing pairs received more feedback than high-performing pairs, indicating that the systems adapted to learners' needs. However, this differentiation was substantially stronger in the reactive condition, where effect sizes for feedback distribution were consistently larger. In the reactive study, low performers received more dialogue prompts and gaze-awareness feedback, while high performers received more task-based hints. This escalation pattern is characteristic of threshold-based systems that respond once misalignment has already occurred. While effective for correcting breakdowns, such designs may inadvertently reinforce performance disparities by intervening more aggressively for struggling pairs.

By contrast, the proactive system exhibited a more balanced feedback ecology. High performers predominantly received "do nothing" feedback, preserving autonomy and flow, while low performers received lighter-weight scaffolds such as dialogue prompts, gaze-awareness feedback, and GitHub Copilot activation. Notably, hint-based feedback no longer differentiated performance levels in the proactive condition. This suggests that anticipatory regulation can reduce reliance on disruptive interventions by maintaining alignment before severe breakdowns occur.

These findings resonate strongly with contemporary views of learner agency as relational and context-sensitive [19, 24, 27]. Rather than positioning learners as passive recipients of corrective feedback, the proactive system supported agency by maintaining learners' involvement in regulatory decisions and minimizing unnecessary intrusion. This aligns with calls to design AI-supported learning environments that enhance, rather than undermine, learners' capacity for shared regulation [4, 15].

### *5.4. Implications for SSRL and Human–AI Collaborative Learning Design*

Collectively, the findings make several contributions to research on SSRL, pair programming, and human–AI collaboration. First, they provide empirical evidence that multimodal indicators such as gaze synchrony and mental-effort synchrony can serve as reliable, actionable triggers for group-level regulation in authentic programming tasks, extending prior observational work in multimodal learning analytics [1, 36, 37].

Second, the results highlight the critical role of feedback timing. While reactive feedback improves performance, proactive feedback reshapes the regulatory landscape by supporting maintenance of optimal states rather than recovery from suboptimal ones. This distinction has important implications for the design of AI agents intended to function as co-regulators of learning.

Finally, the study demonstrates that respecting learner agency, by allowing feedback to be ignored, paused, or implicitly conveyed through "do nothing" signals, does not diminish effectiveness. Instead, it appears to strengthen the partnership between human learners and AI systems, supporting recent arguments for transparent, agency-preserving AI design in education [14, 15].

### *5.5. Limitations and Future Research*

Despite these contributions, the study has limitations. The tasks were conducted in controlled laboratory settings and focused on relatively short debugging activities, which may limit generalizability to longer-term or industrial programming contexts. Additionally, while gaze and mental-effort capture important cognitive-regulatory dimensions, SSRL also involves motivational and affective processes that were not directly measured. Future work should explore longitudinal deployments of proactive SSRL-support systems, integrate additional multimodal signals (e.g., speech or affective cues), and examine how learners' perceptions of AI agency evolve over time. Such research would further clarify how hybrid-intelligent systems can sustainably support socially shared regulation in complex collaborative learning environments.

### *5.6. Contributions*

#### *5.6.1. Contribution 1: Demonstrating the effectiveness of reactive, multimodal SSRL feedback*

The first contribution lies in empirically demonstrating that reactive feedback grounded in joint visual attention and joint mental-effort significantly improves pair programming performance. Study 1 showed that pairs receiving reactive feedback solved more bugs and did so more efficiently than pairs in the control condition, with the strongest effects observed when both JVA- and JME-based feedback were combined.

This contribution directly operationalises and validates SSRL design principles by:

(1) Increasing awareness: Making attentional and cognitive misalignment visible through gaze-awareness feedback.
(2) Supporting externalization: Encouraging partners to verbalize reasoning and resolve asymmetries via dialogue prompts.
(3) Prompting regulation: Triggering task-based hints when sustained cognitive imbalance was detected.

By linking real-time multimodal indicators to these principles, the study advances prior observational work on gaze and collaboration by showing that such indicators can be used not only to analyze SSRL, but also to actively scaffold it.

### 5.6.2. *Contribution 2: Establishing the value of proactive, forecast-based SSRL support*

The second contribution is the demonstration that proactive, anticipatory feedback further enhances collaborative outcomes. Study 2 showed that forecast-based interventions improved debugging success, reduced time on task, and increased feedback uptake compared to a no-feedback condition.

This contribution extends SSRL theory by showing that:

(1) Regulatory support does not need to wait for observable breakdowns to occur.
(2) Forecasted changes in joint attention and cognitive synchrony can serve as effective early triggers for regulation.
(3) Anticipatory feedback supports the maintenance of optimal collaborative states, rather than solely focusing on recovery.

From a design perspective, this contribution illustrates how AI systems can participate in planning and monitoring phases of SSRL, not just in corrective regulation, thereby expanding the functional role of AI in collaborative learning environments.

### 5.6.3. *Contribution 3: Revealing how feedback timing reshapes learner agency and human–AI interaction*

The third contribution concerns how reactive and proactive feedback designs differentially shape human–AI interaction and learner agency. Comparative analyses showed that reactive systems produced stronger differentiation between high- and low-performing pairs, with low performers receiving substantially more, and more intrusive, feedback. In contrast, proactive systems distributed feedback more subtly, allowing high-performing pairs to maintain autonomy while still supporting lower-performing pairs with lightweight scaffolds.

This contribution highlights that:

(1) Feedback timing influences agency: Proactive feedback better preserves learners' sense of control by intervening earlier and less disruptively.
(2) AI can adapt its role dynamically: Acting as a corrective agent in reactive systems and as a maintenance-oriented partner in proactive systems.
(3) "Do nothing" feedback is meaningful: Explicit non-intervention functions as positive regulatory feedback, reinforcing effective collaboration without disruption.

These findings refine design principles for AI-supported SSRL by emphasizing not only what feedback is provided, but when and how it is delivered.

## 6. Conclusions

This paper investigated how feedback grounded in joint visual attention (JVA) and joint mental-effort (JME) can support socially shared regulation of learning (SSRL) in pair programming, and how different feedback logics, reactive versus proactive, shape collaborative performance and human–AI interaction. Across two experimental studies, the findings demonstrate that multimodal, AI-supported feedback systems can substantially improve debugging success, reduce time on task, and influence how learners engage with regulatory support.

Overall, this work contributes empirical evidence, methodological advances, and design insights for the development of human–AI collaborative learning systems that are grounded in SSRL theory and sensitive to learners' cognitive and collaborative states.

## References

[1] Weiqi Xu, Yajuan Wu, and Fan Ouyang. Multimodal learning analytics of collaborative patterns during pair programming in higher education. *International Journal of Educational Technology in Higher Education*, 20(1):8, 2023.

[2] Rolf Steier and Jacob Gorm Davidsen. Adapting interaction analysis to cscl: A systematic review. In *14th International Conference on Computer-Supported Collaborative Learning (CSCL): ISLS Annual Meeting 2021 Reflecting the Past and Embracing the Future*, pages 157–160. International Society of the Learning Sciences (ISLS), 2021.

[3] Yi-Miao Yan, Chuang-Qi Chen, Yang-Bang Hu, and Xin-Dong Ye. Llm-based collaborative programming: impact on students' computational thinking and self-efficacy. *Humanities and Social Sciences Communications*, 12(1):1–12, 2025.

[4] Jinhee Kim, Rita Detrick, Seongryeong Yu, Yukyeong Song, Linda Bol, and Na Li. Socially shared regulation of learning and artificial intelligence: Opportunities to support socially shared regulation. *Education and Information Technologies*, pages 1–39, 2025.

[5] Lauren E Margulieux, James Prather, Brent N Reeves, Brett A Becker, Gozde Cetin Uzun, Dastyni Loksa, Juho Leinonen, and Paul Denny. Self-regulation, self-efficacy, and fear of failure interactions with how novices use llms to solve programming problems. In *Proceedings of the 2024 on Innovation and Technology in Computer Science Education V. 1*, pages 276–282. 2024.

[6] Leonardo Silva, António Mendes, Anabela Gomes, and Gabriel Fortes. Fostering regulatory processes using computational scaffolding. *International Journal of Computer-Supported Collaborative Learning*, 18(1):67–100, 2023.

[7] Cheng-Ye Liu, Wei Li, Ji-Yi Huang, Lu-Yuan Lei, and Pei-Rou Zhang. Collaborative programming based on social shared regulation: An approach to improving students' programming achievements and group metacognition. *Journal of Computer Assisted Learning*, 39(5):1714–1731, 2023.

[8] Chengye Liu, Zhihao Cui, and Xiaojing Weng. Effects of a self-regulated cooperative programming approach on elementary students' programming knowledge, metacognitive awareness, and self-efficacy. *Journal of Computer Assisted Learning*, 42(1):e70167, 2026.

[9] Deller James Ferreira and Dirson Santos de Campos. Investigating how introductory programming students apply regulation strategies. In *CSEDU (2)*, pages 463–473, 2023.

[10] Ahmed Kharrufa, Sami Alghamdi, Abeer Aziz, and Christopher Bull. Llms integration in software engineering team projects: Roles, impact, and a pedagogical design space for ai tools in computing education. *ACM Transactions on Computing Education*, 2024.

[11] Qianou Ma, Tongshuang Wu, and Kenneth Koedinger. Is ai the better programming partner? human-human pair programming vs. human-ai pair programming. *arXiv preprint arXiv:2306.05153*, 2023.

[12] Sanna Järvelä, Andy Nguyen, and Allyson Hadwin. Human and artificial intelligence collaboration for socially shared regulation in learning. *British Journal of Educational Technology*, 54(5):1057–1076, 2023.

[13] Belle Dang, Luna Huynh, Faaiz Gul, Carolyn Ros´e, Sanna Järvelä, and Andy Nguyen. Human–ai collaborative learning in mixed reality: Examining the cognitive and socio-emotional interactions. *British Journal of Educational Technology*, 2025.

[14] Jason M Lodge, Paula de Barba, and Jaclyn Broadbent. Learning with generative artificial intelligence within a network of co-regulation. *Journal of University Teaching and Learning Practice*, 20(7):1–10, 2023.

[15] Justin Edwards, Andy Nguyen, Joni Lämsä, Marta Sobocinski, Ridwan Whitehead, Belle Dang, Anni-Sofia Roberts, and Sanna Järvelä. Human-ai collaboration: Designing artificial agents to facilitate socially shared regulation among learners. *British Journal of Educational Technology*, 56(2):712–733, 2025.

[16] Zipei Ouyang. Self-regulated learning and engagement as serial mediators between ai-driven adaptive learning platform characteristics and educational quality: a psychological mechanism analysis. *Frontiers in Psychology*, 16:1646469, 2025.

[17] Shihui Feng, Huilin Zhang, and Dragan Gaˇsevi´c. Where is aied headed? key topics and emerging frontiers (2020-2024). *arXiv preprint arXiv:2506.20971*, 2025.

[18] Andy Nguyen, Yeyu Wang, Ridwan Whitehead, Muhammad Ashiq, Sanna Järvelä, and David Williamson Shaffer. Examining the interplay of gaze and verbal interactions in socially shared regulation of learning: A transmodal analysis (tma) study. 2025.

[19] Jia He, Haixiao Wang, Jun Xia, and Xiang He. An ecological perspective on learner agency: The case of chinese tertiary-level efl students in peer reviews. *System*, 121:103222, 2024.

[20] Alan Y Cheng, Carolyn Q Zou, Anthony Xie, Matthew Hsu, Felicia Yan, Felicity Huang, David K Zhang, Arjun Sharma, Rashon Poole, Daniel Wan Rosli, et al. Oak story: Improving learner outcomes with llm-mediated interactive narratives. In *Proceedings of the 38th Annual ACM Symposium on User Interface Software and Technology*, pages 1–17, 2025.

[21] Christian Brandmo and Siv M Gamlem. Students' perceptions and outcome of teacher feedback: A systematic review. In *Frontiers in Education*, volume 10, page 1572950. Frontiers Media SA, 2025.

[22] Da Yan. Rubric co-creation to promote quality, interactivity and uptake of peer feedback. *Assessment & Evaluation in Higher Education*, 49(8):1017–1034, 2024.

[23] James Wood and Edd Pitt. Empowering agency through learner-orchestrated self-generated feedback. *Assessment & Evaluation in Higher Education*, 50(1):127–143, 2025.

[24] Juliana Tay, Na Li, Lan Luo, Zihui Zhou, Wan Meng, Qing Zhang, Erick Purwanto, and Yongjia Lu. Role of learner agency for interactive group learning through the lens of blooms taxonomy. *Interactive Learning Environments*, pages 1–14, 2025.

[25] Burhanuddin and Mohammad Arsyad Arrafii. Unfolding the typology and quality of the learner agency practices in the teachers' implementation of the 2013 curriculum in indonesia: the normalisation process theory perspective. *Asia Pacific Education Review*, 24(4):545–561, 2023.

[26] Marion Blumenstein, Asma Shakil, and Peter Swedlund. Technology-enhanced self and peer assessment to support student agency during group projects. *ASCILITE Publications*, 2023.

[27] Hossein Nassaji and Eva Kartchava. Epilogue: The interplay of agency and instruction in written corrective feedback: Reflections and future directions. *Canadian Journal of Applied Linguistics*, 28(1):121–129, 2025.

[28] Pernille Fiskerstrand, Fabienne Van der Kleij, and Wenke Mork Rogne. Editorial students' voices in formative assessment feedback: New insights from research topic contributions. In *Frontiers in Education*, volume 10, page 1698277. Frontiers, 2025.

[29] Jovita Ponomarioviene˙ and Daiva Jakavonyte˙-Staškuvienė. Manifestation of learner agency in primary education: Goal setting, implementation, and reflection in the context of competency-based learning. *Behavioral Sciences*, 15(8):1116, 2025.

[30] Sanna Järvelä, Paul A Kirschner, Ernesto Panadero, Jonna Malmberg, Chris Phielix, Jos Jaspers, Marika Koivuniemi, and Hanna Järvenoja. Enhancing socially shared regulation in collaborative learning groups: Designing for cscl regulation tools. *Educational Technology Research and Development*, 63(1):125–142, 2015.

[31] Leonardo Silva, António Jos´e Mendes, and Anabela Gomes. Computer-supported collaborative learning in programming education: A systematic literature review. In *2020 IEEE Global Engineering Education Conference (EDUCON)*, pages 1086–1095. IEEE, 2020.

[32] Kent Beck and Cynthia Andres. *Extreme Programming Explained: Embrace Change*.

Addison-Wesley Professional, 2004.

[33] Anja Hawlitschek, Sarah Berndt, and Sandra Schulz. Empirical research on pair programming in higher education: a literature review. *Computer science education*, 33(3):400–428, 2023.

[34] David Preston. Pair programming as a model of collaborative learning: A review of the research. *Journal of Computing Sciences in colleges*, 20(4):39–45, 2005.

[35] Norsaremah Salleh, Emilia Mendes, and John Grundy. Empirical studies of pair programming for cs/se teaching in higher education: A systematic literature review. *IEEE Transactions on Software Engineering*, 37(4):509–525, 2010.

[36] Maureen M Villamor and Ma Mercedes T Rodrigo. Gaze collaboration patterns of successful and unsuccessful programming pairs using cross-recurrence quantification analysis. *Research and Practice in Technology Enhanced Learning*, 14(1):25, 2019.

[37] Kshitij Sharma and Giulio Molinari. What happens when collaborators are not in-synch? In *Proceedings of the 16th International Conference on Computer-Supported Collaborative Learning (CSCL 2023)*, pages 99–106, 2023.

[38] Andrew T Duchowski, Krzysztof Krejtz, Cezary Biele, Agnieszka Niedzielska, Peter Kiefer, Martin Raubal, and Ioannis Giannopoulos. The index of pupillary activity: Measuring cognitive load vis-à-vis task difficulty with pupil oscillation. In *Proceedings of the 2018 CHI Conference on Human Factors in Computing Systems*, pages 1–13, 2018.

[39] Bertrand Schneider, Kshitij Sharma, S´everin Cuendet, and Pierre Dillenbourg. Leveraging mobile eye-trackers to capture joint visual attention in co-located collaborative learning groups. *International Journal of Computer-Supported Collaborative Learning*, 13(3):241–261, 2018.

[40] Anahita Golrang and Kshitij Sharma. Does feedback based on gaze and stress indicators help novice programmers? In *European Conference on Technology Enhanced Learning*, pages 198–213. Springer, 2025.

[41] Moreno I Coco and Rick Dale. Cross-recurrence quantification analysis of categorical and continuous time series: an r package. *Frontiers in psychology*, 5:510, 2014.

[42] Eirini Kalliamvakou. Research: Quantifying github copilot's impact on developer productivity and happiness. The GitHub Blog, September 2022. Accessed: 2025-12-02.

## Appendix

**Table 9.:** Feedback combinations for different states

| SCENARIO | JVA | JME | ME1 | ME2 | A1: Do Nothing | A2: GitHub Copilot | A3: Gaze-awareness Tool | A4: Dialog Prompt | A5: Show Hint |
|---|---|---|---|---|---|---|---|---|---|
| 1 | H | H | H | H | | ✓ | | | ✓* |
| 2 | H | H | AVG | AVG | ✓ | | | | |
| 3 | H | H | L | L | | ✓ | | | |
| 4 | H | L | H | H | | ✓ | | ✓ | ✓* |
| 5 | H | L | H | L | | | | ✓ | |
| 6 | H | L | L | L | | | | ✓ | |
| 7 | H | L | AVG | H | | | | ✓ | |
| 8 | H | L | L | L | | | | ✓ | |
| 9 | H | L | L | L | | | | ✓ | |
| 10 | H | L | H | H | | | | ✓ | |
| 11 | H | L | L | L | | ✓ | | ✓ | |
| 12 | L | H | L | L | | ✓ | ✓ | | ✓* |
| 13 | L | H | H | L | | | ✓ | | |
| 14 | L | H | L | L | | ✓ | ✓ | | |
| 15 | L | L | H | H | | ✓ | ✓ | ✓ | ✓* |
| 16 | L | L | H | AVG | | | ✓ | ✓ | |
| 17 | L | L | H | L | | ✓ | ✓ | ✓ | |
| 18 | L | L | AVG | H | | | ✓ | ✓ | |
| 19 | L | L | AVG | L | | | ✓ | ✓ | |
| 20 | L | L | L | H | | ✓ | ✓ | ✓ | |
| 21 | L | L | L | AVG | | | ✓ | ✓ | |
| 22 | L | L | L | L | | ✓ | ✓ | ✓ | |

**Table 10.:** Different types of feedback

| ID | Feedback types |
|---|---|
| A1 | Do nothing (positive feedback) |
| A2 | Offer help through enabling GitHub Copilot |
| A3 | Offer help through enabling the gaze-awareness tool |
| A4 | Prompt to initiate dialogue |
| A5 | Prompt a task-based hint |

## Appendix A. Detailed Scenario Logic and Feedback Combinations

### *A.1. Overview*

Table 9 defines 22 scenarios derived from combinations of Joint Visual Attention (JVA), Joint mental-effort (JME), and individual mental-efforts ($ME_1$, $ME_2$). Each configuration reflects a distinct collaboration state and triggers specific feedback mechanisms (**A1–A7**) in the Oculii system. Desired states correspond to high JVA and JME with average individual MEs, whereas suboptimal configurations invoke corrective or assistive feedback.

### *A.2. Scenario Descriptions*

- **High JVA – High JME (Scenarios 1–3):** Optimal engagement states; use of A1 (“do nothing”) or A2/A5 for cognitive rebalancing.
- **High JVA – Low JME (Scenarios 4–11):** Visual alignment but cognitive divergence; initiate A4 (dialog) and optionally A2/A5.
- **Low JVA – High JME (Scenarios 12–14):** Productive division of labour; maintain A3 (gaze-awareness) and support with A2 as needed.
- **Low JVA – Low JME (Scenarios 15–22):** Disengaged or uncoordinated states; combine A3, A4, and when required A2/A5 to restore synchrony.

### *A.3. Lookup Table of Feedback Combinations*

### *A.4. Summary*

By categorizing parameter configurations into discrete scenarios, the system modulates between non-disruptive feedback (**A1–A3**) that maintains workflow and more disruptive prompts (**A4–A6**) that restore cognitive balance. The final feedback type, **A7**, remains user-initiated, ensuring transparency and user control over the adaptive process.

### *A.5. Scenario-Based Feedback Logic*

To operationalize adaptive feedback, the system categorizes combinations of **Joint Visual Attention (JVA)**, **Joint mental-effort (JME)**, and **individual mental-effort (ME)** levels into discrete collaboration scenarios (see Table 9). Each scenario represents a specific cognitive–collaborative configuration of the pair, allowing the system to trigger appropriate feedback types (**A1–A7**) to maintain or restore an optimal collaboration state. Desired states are characterized by *high JVA*, *high JME*, and *average individual MEs*—conditions under which partners are jointly attentive, mutually engaged, and cognitively balanced.

#### *A.5.1. High JVA – High JME (Scenarios 1–3)*

These scenarios indicate strong shared attention and engagement. However, only **Scenario 2** constitutes a *desired state*, where both individual MEs are balanced. Here, the system applies **A1 (Do nothing)** to preserve the optimal configuration.

In **Scenarios 1** and **3**, the individual MEs are both too high or too low, respectively. To guide the pair toward the optimal average ME level, **A2 (GitHub Copilot)** is activated—either to aid in discovering new solutions under excessive cognitive load or to accelerate task progress when the task is too easy. If the high-ME condition persists (*Scenario 1* ), a corrective hint **(A5)** is subsequently provided to re-stabilize the pair's cognitive alignment.

#### *A.5.2. High JVA – Low JME (Scenarios 4–11)*

Here, both partners remain visually coordinated, but their cognitive engagement diverges. The system prioritizes **A4 (Dialog Prompt)** to stimulate discussion and restore shared understanding. Dialogic reflection can help align individual effort levels, transforming low JME into a higher, more synchronized state.

When both MEs are extreme—either high (*Scenario 4* ) or low (*Scenario 11* )—additional aids such as **A2 (Copilot)** and, where necessary, **A5 (Hint)** are enabled to regulate cognitive load and re-establish balance across the pair.

#### *A.5.3. Low JVA – High JME (Scenarios 12–14)*

These scenarios represent *desired states* of effective Division of Labour (DoL)—either task-based or role-based. Although the pair is productively engaged, low JVA may hinder coordination. To support smooth transitions between sub-tasks, **A3 (Gaze-awareness tool)** is activated, allowing partners to re-synchronize their visual focus when needed.

In **Scenarios 12** and **14**, **A2 (Copilot)** aids code exploration or rapid navigation. For **Scenario 14**, which may indicate either deep focus or disengagement, **A6 (Question Prompt)** is occasionally employed to assess the collaboration state and help re-engage the pair when necessary.

#### *A.5.4. Low JVA – Low JME (Scenarios 15–22)*

Low levels of both indicators suggest disconnection or reduced engagement. Consequently, **A3 (Gaze-awareness)** and **A4 (Dialog Prompt)** are deployed to rebuild mutual focus and stimulate discussion. For configurations where both individual MEs deviate from average (*Scenarios 15, 17, 20, 22* ), **A2 (Copilot)** provides external scaffolding to encourage progress. In **Scenario 15**, sustained high MEs trigger an additional **A5 (Hint)** to alleviate overload.

#### *A.5.5. Summary and System Behavior*

By mapping the multimodal parameters to scenario clusters, the system balances *non-disruptive feedback* (**A1–A3**) that preserves task flow with *disruptive interventions* (**A4–A6**) that drive cognitive realignment when collaboration deteriorates. Addition-ally, **A7**, which is user-initiated through the interface, provides transparency and user agency by offering explanations of the feedback logic and system behavior—an important element for maintaining user trust and perceived control.